# Beyond epistemological deficits:  Incorporating flexible epistemological views into fine-grained cognitive dynamics

Ayush Gupta, Andy Elby, Department of Physics, University of Maryland, College Park, MD 20742.
Email: ayush@umd.edu, elby@umd.edu

**Abstract:** Researchers have argued against deficit-based explanations of students' troubles with mathematical sense-making, pointing instead to factors such as epistemology: students' beliefs about the nature of knowledge and learning can hinder them from activating and integrating productive knowledge they have.  But such explanations run the risk of substituting an epistemological deficit for a concepts/skills deficit.  Our analysis of an undergraduate engineering major avoids this "deficit trap" by incorporating multiple, context-dependent epistemological stances into his cognitive dynamics.

## Introduction

Engineering students often have trouble learning and using mathematics effectively (Redish & Smith, 2008), and epistemologies play a major role in explaining these difficulties (Hammer, 1994; Schoenfeld, 1988; Schommer, Crouse, & Rhodes, 1992).  Our case study of a student who gets stuck while solving a problem during a clinical interview adds to this literature in two ways. First, we provide a fine-grained account of *how* Jim's epistemology affects his mathematical problem-solving in the moment.  This complements large-*N* studies showing that epistemological beliefs correlate with approaches to learning and using mathematics and with performance (reviewed by Muis, 2004).  Second, we show that Jim's "problematic" epistemological stance is nuanced and flexible in response to conceptual cues, and we provide a toy model that can explain his subtle shift between epistemological stances.  In this way, we avoid attributing to Jim an epistemological deficit, i.e., an absence of productive epistemological resources.  Our analysis also has instructional implications:  instead of needing to confront and replace "mis-epistemologies" such as the one Jim initially appears to hold, instructors can create instructional environments that tend to trigger more favorable epistemological stances.

## Methods and theoretical framework

We videotaped clinical interviews of Jim and six other engineering majors taking a first-semester physics course, probing their approaches to using mathematics in physics.  Subjects were asked to explain both a familiar and an unfamiliar equation to themselves and to others, and to solve problems while thinking aloud. Jim solved one problem easily, but got stuck on another despite his fluidity with mathematical manipulation and his understanding of the relevant physical ideas.  We thought that understanding why Jim got stuck — and then, why a particular conceptual cue helped him get past the impasse — could shed light on the cognitive dynamics by which epistemological stances and conceptual ideas interact during mathematical problem-solving.

     Our analysis began with close scrutiny of the episode where Jim gets stuck and later unstuck, borrowing tools from discourse analysis (Gee, 1999) and framing analysis (Tannen, 1993) to interpret gestures, word choice, and the contextualized substance of his utterances, taking into account that social expectations and power dynamics play a role in interviews. We formed explanations for his behavior and looked for confirmatory or disconfirmatory evidence elsewhere in the interview (Miles & Huberman, 1984). Working from a knowledge-in-pieces perspective (diSessa, 1993; Minsky, 1986), we did not attribute patterns of behavior to globally robust (mis)conceptions and epistemological beliefs; we continually considered how contextual cues might trigger different local coherences in his thinking (Hammer, Elby, Scherr, & Redish, 2005).

## Data and analysis:  Brief synopsis

The interviewer asks whether the pressure 7 meters beneath the surface of a lake is greater than or less than pressure 5 meters beneath the surface.  Jim uses the equation $p = p_{at\ top} + \rho gh$, to which he had just been introduced ($p$, $\rho$, and $g$ denote, pressure, density of water, and acceleration due to gravity, and $h$ denotes the distance beneath the surface. Jim thinks $h$ must be negative (i.e., 5 meters beneath the surface corresponds to $h = -5$ m), and concludes that the $p$ is greater at a depth of 5 meters. The interviewer tries to correct the sign error:

> Interviewer: Suppose I told you that $h$ is positive.
> Jim:      Always positive?
> Int.:       Yes, ... Would that help you?
> Jim:      I mean, that would just make 7 greater than 5.
> Int.:      Okay. Does that bother you?
> Jim:      I mean... What I keep thinking is that you are going *down* (gestures down) so 7 cannot be greater than 5 and negative. That's why I keep coming back to that.

> Meaning, if you do say it's positive then ... I guess it doesn't bother me. (sighs) 7 is greater than 5 in positive-land.

Jim is skeptical about $h$ being is positive, given that its direction is *down* from the surface. Jim sighs as he tries to accept that idea and perhaps tries to distance himself from it by saying it's true in "positive-land," which he says with a hint of sarcasm. In the conversation that follows, Jim reverts to thinking of $h$ as negative.

The interviewer then asks Jim how a friend who doesn't know physics would answer this question. Jim says that, based on personal experiences, the friend would say the pressure is higher at greater depths. Jim acknowledges the conflict between the common-sense answer and the mathematical answer, but thinks the mathematical answer is more trustworthy partly because perceptions can be misleading:

> For an equation to be given to you it has to be like theory and it has to be fact-bearing. So, fact applies for everything. It is like a law. It applies to every single situation you could be in. But, like, your experience at times or perception is just different - or you don't have the knowledge of that course or anything. So, I will go with the people who have done the law and it has worked time after time after time.

He doesn't try to reconcile "perceptions" with math, though he has the tools to do so, as shown below.

At first glance, we might think Jim's behavior stems from the robust conception that downward always corresponds to negative. But solving an earlier problem, Jim took downward as positive. We can also rule out conceptual or skill deficits, using data we lack space to present here. The reason he gets and stays stuck, we argue, is that his perception of how difficult it would be to reconcile common sense and the formal result (while he's "on the spot" in the interview!) stabilizes his epistemological stance that mathematical formalism and common-sense ideas need not speak to each other, and that mathematical formalism expresses confirmed truths.

This stance is more nuanced and flexible than a globally-robust "belief" that mathematical formalism is disconnected from everyday thinking or the real world. When the interviewer asks about the sign of $g$, Jim immediately reconciles the mathematics with common sense: with $g$ and $h$ both taken as negative, $\rho g h$ is positive and hence the pressure is greater 7 meters beneath the surface. Upon reaching this conclusion, Jim displays relief in his physical posture and his vocal pitch while noting that this makes more sense and agrees with the physical experience of pressure. We argue that his epistemological stance has shifted in this moment (not necessarily permanently!), towards a greater expectation that mathematical formalism should mesh with everyday experience. And this shift is actually quite subtle: even before the reconciliation, in response to the interviewer's query about whether the mathematical formalism relates to real-life experiences under water, he adds, "I think there is some way that just completely links the two together, but it's not obvious what that relation is." So, Jim always thought a reconciliation was possible, but he thought achieving it would be prohibitively difficult until the interviewer pointed him toward a path. In our poster, we will model Jim's stances before and after the reconciliation as networks of finer-grained knowledge elements, and the subtle *epistemological* shift as a cascade resulting from a *conceptual* cue.